\documentclass[aps,twocolumn,showpacs]{revtex4}
\usepackage{epsfig}
\usepackage{graphicx}
\usepackage{amsfonts}
\usepackage[figuresright]{rotating}
\usepackage{amssymb}
\usepackage{amsmath}
\usepackage{psfrag}

\def\avg#1{\langle#1\rangle}

\def\be{\begin{equation}} \def\ee{\end{equation}}
\def\bea{\begin{eqnarray}} \def\eea{\end{eqnarray}}

\def\pp{\parallel}

\begin{document}

\title{The $J$-triplet Cooper pairing with magnetic dipolar interactions}

\author{Yi Li}
\affiliation{Department of Physics, University of California, San Diego,
CA 92093 }

\author{Congjun Wu}
\affiliation{Department of Physics, University of California, San Diego,
CA 92093 }

\begin{abstract}

Recently, cold atomic Fermi gases with the large magnetic dipolar interaction
have been laser cooled down to quantum degeneracy.
Different from electric-dipoles which are classic vectors, atomic magnetic
dipoles are quantum-mechanical matrix operators proportional to the
hyperfine-spin of atoms, thus provide rich opportunities to investigate
exotic many-body physics.
Furthermore, unlike anisotropic electric dipolar gases, unpolarized magnetic
dipolar systems are isotropic under simultaneous spin-orbit rotation.
These features give rise to a robust mechanism for a novel pairing
symmetry: orbital $p$-wave ($L=1$) spin triplet ($S=1$) pairing with total angular
momentum of the Cooper pair $J=1$.
This pairing is markedly different from both the $^3$He-B phase in which
$J=0$ and the $^3$He-$A$ phase in which $J$ is not conserved.
It is also different from the $p$-wave pairing in the single-component
electric dipolar systems in which the spin degree of freedom is frozen.
\end{abstract}


\maketitle

Ultracold atomic and molecular systems with electric and magnetic dipolar
interactions have become the research focus in cold atom physics
\cite{ospelkaus2008,ni2008,griesmaier2005,mcclelland2006,lu2010,youn2010, Aikawa2012}.
When dipole moments are aligned by external fields, dipolar
interactions exhibit the $d_{r^2-3z^2}$-type anisotropy.
The anisotropic Bose-Einstein condensations of dipolar bosons (e.g. $^{52}$Cr)
have been observed \cite{koch2008,lahaye2009,lahaye2009a,menotti2007a}.
For the fermionic electric dipolar systems, $^{40}$K-$^{87}$Rb has been
cooled down to nearly quantum-degeneracy \cite{ospelkaus2008}.
Effects of the anisotropic electric dipolar interaction on
the fermion many-body physics have been extensively investigated.
In the Fermi liquid theory, both the single particle properties and
collective excitations exhibit the $d_{r^2-3z^2}$ anisotropy
\cite{sogo2009,miyakawa2008,fregoso2009,ronen2010,chan2010,lin2010}.
In the single-component Fermi systems, the leading order Cooper pairing
instability lies in the $p$-wave channel, which is the simplest one allowed
by Pauli's exclusion principle.
The anisotropy of the electric dipolar interaction selects the instability
in the $p_z$-channel, which is slightly hybridized with other odd partial
wave channels \cite{baranov2002,baranov2004,baranov2008a,you1999,bruun2008,
levinsen2011,potter2010,lutchyn2010}.
For two-component cases, the dipolar interaction leads to
anisotropic spin-triplet pairing, and its orbital
partial wave is again in the $p_z$-channel \cite{samokhin2006,
wu2010,shi2009,kain2011}.
The triplet pairing competes with the singlet pairing in the hybridized
$s+d_{r^2-3z^2}$-channel.
The mixing between the singlet and triplet pairings has a relative
phase $\pm \frac{\pi}{2}$, which leads to a novel time-reversal symmetry
breaking Cooper pairing state \cite{wu2010}.

An important recent experimental progress is the laser cooling and trapping
of magnetic dipolar fermions of $^{161}$Dy and $^{163}$Dy with large
atomic magnetic moments $(10\mu_B)$ \cite{lu2010,youn2010}.
There are important differences between magnetic and electric dipolar
interactions.
Electric dipole moments are essentially non-quantized classic vectors
from the mixing between different rotational eigenstates,
which are
induced by external electric fields \cite{ospelkaus2008,ni2008}, thus
electric dipoles are frozen.
In the absence of external fields, even though at each instant of time
there is a dipole moment of the heteronuclear molecule, it is averaged
to zero at a long time scale.
In contrast, magnetic dipole moments of atoms are intrinsic, proportional to
their hyper-fine spins with a Lande factor.
Unpolarized magnetic dipolar Fermi systems are available, in which dipoles are
defrozen as non-commutative quantum mechanical operators, thus lead to
richer quantum spin physics of dipolar interactions.
Furthermore, the magnetic dipolar interaction is actually isotropic in the
unpolarized systems.
It is invariant under simultaneous spin-orbit rotations but not separate
spin or orbit rotations.
This spin-orbit coupling is different from usual single particle one,
but an interaction effect.
It plays an important role in the Fermi liquid properties such as
the unconventional magnetic states and ferro-nematic states
predicted by Fregoso {\it et al} \cite{fregoso2010,fregoso2009a}.

It is natural to expect that magnetic dipolar interaction brings
novel pairing symmetries not studied in condensed matter systems before.
The systems of $^{161}$Dy and $^{163}$Dy are with a very large hyperfine
spin of $F=\frac{21}{2}$, thus their Cooper pairing problem is expected
to be very challenging.
As a first step, we study the simplest case of spin-$\frac{1}{2}$,
and find that the magnetic dipolar interaction provides a novel and robust
mechanism to the $p$-wave ($L=1$) spin triplet ($S=1$) Cooper pairing to
the first order of interaction strength, which comes from the attractive part
of the magnetic dipolar interaction.
In comparison, the $p$-wave triplet pairing in usual condensed matter systems,
such as $^3$He \cite{anderson1961,balian1963,brinkman1974}, is due to the
spin-fluctuation mechanism, which is at the second order of interaction strength
(see Refs. [\onlinecite{leggett1975,volovik2009}] for reviews).
This mechanism is based on strong ferromagnetic tendency from the
repulsive part of the $^3$He-$^3$He interactions.
Furthermore, the $p$-wave triplet Cooper pairing symmetry patterns
in magnetic dipolar systems are novel, which do not appear in $^3$He.
The orbital and spin angular momenta of the Cooper pair are entangled into
the total angular momentum $J=1$, which is denoted as the $J$-triplet channel
below.
In contrast, in the $^3$He-$B$ phase \cite{balian1963}, $L$ and $S$ are combined
into $J=0$; and in the  $^3$He-$A$ phase, $L$ and $S$ are decoupled
and $J$ is not well-defined \cite{brinkman1974,anderson1961}.
There are two competing pairing possibilities in this $J$-triplet channel
with different values of $J_z$: the helical polar state ($J_z=0$)
preserving time reversal (TR) symmetry, and the axial state ($J_z=\pm 1$)
breaking TR symmetry.
The helical polar state has point nodes and gapless Dirac spectra,
which is a time-reversal invariant generalization of the $^3$He-$A$ phase
with entangled spin and orbital degrees of freedom.
In addition to usual phonon modes, its Goldstone modes contain the total
angular momentum wave as entangled spin-orbital modes.

We begin with the magnetic dipolar interaction between spin-$\frac{1}{2}$
fermions
\bea
V_{\alpha\beta,\beta^\prime\alpha^\prime}(\vec r)= \frac{\mu^2}{r^3}
\big\{\vec S_{\alpha\alpha^\prime}
\cdot \vec S_{\beta\beta^\prime}-3 (\vec S_{\alpha\alpha^\prime} \cdot \hat r)
(\vec S_{\beta\beta^\prime} \cdot \hat r) \big \},
\label{eq:mgdplr}
\eea
where $\vec r$ is the relative displacement vector between two fermions;
$\mu$ is the magnitude of the magnetic moment.
Such an interaction is invariant under the combined
$SU(2)$ spin rotation and $SO(3)$ space rotation.
In other words, orbital angular momentum $\vec L$ and spin $\vec S$ are not
separately conserved, but the total angular momentum $\vec J=
\vec L+\vec S$ remains conserved.
Its Fourier transformation reads \cite{fregoso2009a}
\bea
V_{\alpha\beta;\beta^\prime\alpha^\prime}(\vec q)=\frac{4\pi}{3}
\mu^2 \big\{ 3 (\vec S_{\alpha\alpha^\prime} \cdot \hat q)
(\vec S_{\beta\beta^\prime} \cdot \hat q)-\vec S_{\alpha\alpha^\prime}
\cdot \vec S_{\beta\beta^\prime}\big\}.
\eea
The Hamiltonian in the second quantization form is written as
\bea
H&=& \sum_{\vec k,\alpha} \big[\epsilon(\vec k)-\mu_c \big ]
c^\dagger_{\alpha}(\vec k) c_{\alpha} (\vec k)
+\frac{1}{2V} \times \notag \\
&& \sum_{\vec k,\vec k^\prime,\vec q}
V_{\alpha\beta;\beta^\prime\alpha^\prime}(\vec k-\vec k^\prime)
P^\dagger_{\alpha\beta} (\vec k;\vec q)  P_{\beta^\prime\alpha^\prime}
(\vec k^\prime;\vec q),
\label{eq:ham}
\eea
where $\epsilon(\vec k)=\hbar k^2/(2m)$; $\mu_c$ is the chemical
potential; $P_{\beta^\prime\alpha^\prime} (\vec k;\vec q)
=c_{\beta^\prime}(-\vec k+\vec q) c_{\alpha^\prime}(\vec k+\vec q)$ is
the pairing operator; the Greek indices $\alpha,\beta,\alpha^\prime$
and $\beta^\prime$ refer to $\uparrow$ and $\downarrow$;
$V$ is the volume of the system.
We define a dimensionless parameter characterizing the interaction strength
as the ratio between the characteristic interaction energy and the
Fermi energy: $\lambda \equiv E_{int}/E_F = \frac{2}{3}
\frac{\mu^2 m k_f }{\pi^2\hbar^2}$.

We next study the symmetry of the Cooper pairing in the presence
of Fermi surface, i.e., in the weak coupling theory.
An important feature of the magnetic dipolar interaction in
Eq. (\ref{eq:mgdplr}) is that it vanishes
in the total spin singlet channel. Thus, we only need to study the
triplet pairing in odd orbital partial wave channels.
Considering uniform pairing states at the mean-field level, we
set $\vec q=0$ in Eq. (\ref{eq:ham}), and define triplet pairing
operators $P_s(\vec k)$, which are eigen-operators of
$\vec S_{1z}+\vec S_{2z}$ with eigenvalues $s_z=0,\pm 1$, respectively.
More explicitly, they are
$P_0(\vec k)=\frac{1}{\sqrt 2} [P_{\uparrow\downarrow}(\vec k)
+P_{\downarrow\uparrow}(\vec k)],$
$P_1(\vec k)=P_{\uparrow\uparrow}(\vec k)$,
$P_{-1}(\vec k)=P_{\downarrow\downarrow}(\vec k)$.
The pairing interaction of Eq. (\ref{eq:ham}) reduces to
\bea
H_{pair}&=& \frac{1}{2V}\sum_{\vec k,\vec k^\prime,s_z s_z^\prime}
\Big\{ V^T_{s_zs_z^\prime}(\vec k;\vec k^\prime)
P^{\dagger}_{s_z}(\vec k) P_{s_z^\prime}(\vec k^\prime) \Big\},
\label{Eq:polarpair}
\eea
where
\bea
V^T_{s_zs_z^\prime}(\vec k;\vec k^\prime)&=&
\frac{1}{2}\sum_{\alpha\beta\beta^\prime\alpha^\prime}
\langle 1s_z|\frac{1}{2} \alpha \frac{1}{2}\beta\rangle
\langle 1s_z^\prime|\frac{1}{2} \alpha^\prime \frac{1}{2}\beta^\prime\rangle^*
\notag \\
&& 
\big\{ V_{\alpha\beta,\beta^\prime\alpha^\prime}(\vec k-\vec k^\prime)
-V_{\alpha\beta,\beta^\prime\alpha^\prime}(\vec k+\vec k^\prime)\big\}. \ \ \ \ \ \
\eea
$\langle 1s_z|\frac{1}{2}\alpha\frac{1}{2} \beta\rangle$
is the Clebsch-Gordan coefficient for two spin-$\frac{1}{2}$
states to form the spin triplet; and
$V_{s_zs_z^\prime}(\vec k;\vec k^\prime)$ is an odd function of both
$\vec k$ and $\vec k^\prime$.

The decoupled mean-field Hamiltonian reads
\bea
H_{mf}&=&\frac{1}{2V}\sum_{\vec k} ~^\prime ~ \Psi^\dagger (\vec k)
\left(\begin{array}{cc}
\xi(\vec k) I & \Delta_{\alpha\beta} (\vec k) \\
\Delta^*_{\beta\alpha}(\vec k) &-\xi(\vec k) I
\end{array}
\right) \Psi(\vec k), \ \ \
\label{eq:mf}
\eea
where we only sum over half of the momentum space;
$\xi (\vec k)=\epsilon (\vec k) -\mu_{ch}$ and $\mu_{ch}$
is the chemical potential;
$\Psi(\vec k)=(c_\uparrow (\vec k), c_\downarrow(\vec k),
c_\uparrow^\dagger(-\vec k), c_\downarrow^\dagger(-\vec k) )^T$;
$\Delta_{\alpha\beta}$ is defined as
$\Delta_{\alpha\beta}=\sum_{s_z}\langle 1s_z
|\frac{1}{2}\alpha\frac{1}{2}\beta\rangle^*
\Delta_{s_z}$.
$\Delta_{s_z}$ satisfies the mean-field gap function as
\bea
&&\Delta_{s_z}(\vec k)=\frac{1}{V} \sum_{\vec k^\prime,s_z^\prime}
V^T_{s_zs_z^\prime}(\vec k;\vec k^\prime) \avg{|P_{s_z^\prime} (\vec k^\prime)|}
\notag \\
&=&-\int \frac{d^3 k^\prime}{(2\pi)^3}
V^T_{s_zs_z^\prime}(\vec k;\vec k^\prime) [K(\vec k^\prime)-\frac{1}{2\epsilon_k}]
\Delta_{s^\prime_z} (\vec k^\prime), \ \ \,
\label{eq:gap}
\eea
where $K(\vec k^\prime)=\tanh[\frac{\beta}{2}
E_{i}(\vec k^\prime)]/[2E_i (\vec k^\prime)]$.
The integral in Eq. (\ref{eq:gap}) is already normalized
following the standard procedure \cite{baranov2002}.
For simplicity, we use the Born approximation in Eq. (\ref{eq:gap}) by
employing the bare interaction potential rather than the fully
renormalized $T$-matrix,
which applies in the dilute limit of weak interactions.
The pairing symmetry, on which we are interested below, does not
depend on the details that how the integral of Eq. (\ref{eq:gap}) is
regularized in momentum space.
The Bogoliubov quasiparticle spectra become
$E_{1,2}(\vec k)=\sqrt{\xi_k^2+\lambda^2_{1,2}(\vec k) }$, where
$\lambda^2_{1,2}(\vec k)$ are the eigenvalues of the positive-definite
Hermitian matrix $\Delta^\dagger(\vec k) \Delta(\vec k)$.
The free energy can be calculated as
\bea
F&=&-\frac{2}{\beta}\sum_{\vec k,i=1,2} \ln \big[2\cosh
\frac{\beta E_{\vec k,i} }{2}\big]
\notag \\
&&
-\frac{1}{2V} \sum_{\vec k,\vec k^\prime,s_z,s_z^\prime}
 \big \{ \Delta^*_{s_z}(\vec k) V^{T,-1}_{s_z s_z^\prime}(\vec k; \vec k^\prime)
\Delta_{s_z^\prime}(\vec k)
\big\},
\label{eq:free}
\eea
where $V^{T,-1}_{s_z s_z^\prime}(\vec k; \vec k^\prime)$ is the inverse
of the interaction matrix defined as
\bea
\frac{1}{V}\sum_{\vec k^\prime, s_z^\prime} V^T_{s_z,s_z^\prime}(\vec k; \vec k^\prime)
V^{T,-1}_{s_z^\prime,s_z^{\prime\prime}}(\vec k^\prime; \vec k^{\prime\prime})
=\delta_{\vec k, \vec k^{\prime\prime}} \delta_{s_z,s_z^{\prime\prime}}.
\eea

We next linearize Eq. (\ref{eq:gap}) around $T_c$ and perform the
partial wave analysis to determine the dominant pairing channel.
Since the total angular momentum is conserved, we can use
$J$ to classify the eigen-gap functions denoted as
$\phi^{a,JJ_z}_{s_z} (\vec k)$.
The index $a$ is used to distinguish different channels sharing
the same value of $J$.
$\phi^{a,JJ_z}_{s_z} (\vec k)$ satisfies
\bea
N_0 \int \frac{d\Omega_{k^\prime}}{4\pi} V^T_{s_z s_z^\prime} (\vec k; \vec k^\prime)
\phi^{a;JJ_z}_{s_z^\prime} (\vec k^\prime)
=w^a_{J} \phi^{a;JJ_z}_{s_z}(\vec k),
\eea
where $N_0=\frac{mk_f}{\pi^2\hbar^2}$ is the density of state at
the Fermi surface; $w^a_j$ are dimensionless eigenvalues;
$\vec k,\vec k^\prime$ are at the Fermi surface.
Then Eq. (\ref{eq:gap}) is linearized into a set of decoupled equations
\bea
\phi^{a;JJ_z} \{ 1 +w^a_J [\ln (2 e^{\gamma} \bar \omega)/(\pi k_B T)] \}=0,
\label{eq:Tc}
\eea
where $\bar\omega$ is an energy scale at the order of the Fermi energy
playing the role of energy cut-off from the Fermi surface.

The decomposition of  $V^T_{s_zs_z^\prime}
(\vec k;\vec k^\prime)$ into spherical harmonics can be formulated as
\bea
&&\frac{N_0}{4\pi} V^T_{s_zs_z^\prime}(\vec k;\vec k^\prime)
\notag \\
&=&\sum_{Lm,L^\prime m^\prime}
V_{Lms_z;L^\prime m^\prime s_z^\prime}
Y_{Lm}^*(\Omega_k) Y_{L^\prime m^\prime} (\Omega_{\vec k^\prime}),
\label{eq:partialwave}
\eea
where $L=L^\prime$ or $L=L^\prime\pm2$, and $L, L^\prime$ are odd
numbers.
The expressions of the dimensionless matrix elements
$V_{Lms_z;L^\prime m^\prime s_z^\prime}$ are lengthy
and will be presented elsewhere.
By diagonalizing this matrix, we find that the most negative eigenvalue
is $w^{J=1}=- 3\pi\lambda/4$ lying in the channel with $J=L=1$.
All other negative eigenvalues are significantly smaller.
Therefore, dominate pairing symmetry is identified as the $J$-triplet channel
with $L=S=1$ in the weak coupling theory.
Following the standard method in Ref. \cite{baranov2002}, the transition
temperature $T_c$ is expressed as $T_c\approx \frac{2e^\gamma \bar \omega}
{\pi} e^{-\frac{1}{|w^{J=1}|}}$.
For a rough estimation of the order of magnitude of $T_c$, we set
the prefactor in the expression of $T_c$ as $E_f$.

In order to understand why the $J$-triplet channel is selected by the magnetic
dipolar interaction, we present a heuristic picture based on a two-body
pairing problem in real space.
Dipolar interaction has a characteristic length scale $a_{dp}=m \mu^2/\hbar^2$
at which the kinetic energy scale equals the interaction energy scale.
We are not interested in solving the radial equation but
focus on the symmetry properties of the angular solution, thus,
the distance between two spins is taken fixed at $a_{dp}$.
We consider the lowest partial-wave, $p$-wave, channel with $L=1$.
The $3\times 3=9$ states $(L=S=1)$ are classified into three
sectors of $J=0,1$ and $2$.
In each channel of $J$, the interaction energies are diagonalized as
\bea
E_0=E_{dp}, \ \ \, E_1=-\frac{1}{2} E_{dp}, \ \ \,  E_2=\frac{1}{10} E_{dp},
\eea
respectively,
where $E_{dp}=\mu^2/a^3_{dp}$.
Only the total angular momentum triplet sector with $J=1$ supports bound
states, thus is the dominant pairing channel and is consistent
with the pairing symmetry in the weak-coupling theory.

\begin{figure}
\centering\epsfig{file=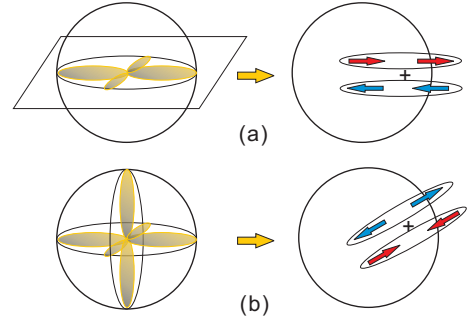,clip=1,width=0.7\linewidth, angle=0}
\caption{The spin configurations of the two-body states
with a) $J=1$ and $j_z=0$ and b) $J=j_z=0$.
The interactions are attractive in a) but repulsive in b).
}
\label{fig:twbdy}
\end{figure}

This two-body picture applies in the strong coupling limit.
Although a complete study of the strong coupling problem is beyond
the scope of this paper, this result provides
an intuitive picture to understand pairing symmetry in the $J$-triplet
sector from spin configurations.
We define that $\chi_\mu$ and $p_\mu(\hat\Omega)$ are eigenstates with
eigenvalues zero for operators $\hat e_\mu \cdot (\vec S_1 +\vec S_2)$
and $\hat e_\mu \cdot \vec L$ $(\mu=x,y,z)$, which are the total spin
and orbital angular momenta projected along the $e_\mu$-direction.
The $J$-triplet sector states
are $\phi_\mu(\Omega)=\frac{1}{\sqrt{2}}
\epsilon_{\mu\nu\lambda} \chi_\nu p_\lambda(\Omega)$ with $\phi_\mu$
satisfying $(\hat e_\mu \cdot \vec J) \phi_\mu=0$.
For example,
\bea
\phi_z(\hat \Omega)&=&
\frac{1}{\sqrt 2}[\chi_x p_y(\hat \Omega)-\chi_y p_x (\hat \Omega) ]
\notag \\
&=&\sqrt{\frac{3}{2}}\sin\theta \big\{|\alpha_{\hat e_\rho} \rangle_1
|\alpha_{\hat e_\rho} \rangle_2+|\beta_{\hat e_\rho} \rangle_1
|\beta_{\hat e_\rho}\rangle_2 \big\},
\eea
where $\hat e_\rho=\hat x \cos\phi+\hat y \sin \phi $
and $|\alpha_{e_\rho}\rangle$ and $|\beta_{e_\rho}\rangle$ are eigenstates of
$\hat e_\rho \cdot \vec \sigma $ with eigenvalues of $\pm 1$.
As depicted in Fig. \ref{fig:twbdy} (a), along the equator where $\phi_z$ has the largest
weight, two spins are parallel and along $\hat r$, thus
the interaction is dominated by attraction.
On the other hand, the eigenstate of $J=0$ reads
\bea
\phi_0(\Omega)=\chi_\mu p_\mu(\Omega)
=\frac{1}{\sqrt 2} \big\{|\alpha_\Omega \rangle_1 |\beta_\Omega\rangle_2+
|\beta_\Omega \rangle_1 |\alpha_\Omega\rangle_2 \big\},
\eea
where $|\alpha_\Omega\rangle$ and $|\beta_\Omega\rangle$ are eigenstates
of $\hat \Omega \cdot \vec \sigma$ with eigenvalues $\pm 1$.
As shown in Fig. \ref{fig:twbdy}
(b), along any direction of $\hat \Omega$, two spins are anti-parallel
and longitudinal, thus the interaction is repulsive.

Let us come back to momentum space and study the competition between
three paring branches in the $J$-triplet channel under the
Ginzburg-Landau (GL) framework.
We define
\bea
\Delta_x(\vec k)&=&\frac{1}{\sqrt 2}[-\Delta_1(\vec k)
+\Delta_{-1} (\vec k)], \notag \\
\Delta_y(\vec k)&=&\frac{i}{\sqrt 2}
[\Delta_1(\vec k)+\Delta_{-1}(\vec k)], \notag \\
\Delta_z(\vec k)&=&\Delta_0(\vec k).
\eea
The bulk pairing order parameters are defined as $\Delta_\mu=
\frac{1}{V}\sum_k \hat k_\mu \Delta_\mu (\vec k)$, where
no summation over $\mu$ is assumed.
We define pairing parameters and their real and imaginary
parts as the following 3-vectors
$\vec \Delta=(\Delta_x,\Delta_y,\Delta_z)$.
The GL free energy is constructed to maintain the $U(1)$
and $SO(3)$ rotational symmetry as
\bea
F&=& \alpha \vec \Delta^* \cdot \vec \Delta
+\gamma_1  |\vec \Delta^* \cdot \vec \Delta |^2
+
\gamma_2  |\vec \Delta^* \times \vec \Delta|^2,
\eea
where
\bea
\alpha=N_0 \ln (\frac{T}{T_c}).
\label{eq:Tc}
\eea
The sign of $\gamma_2$ determines two different pairing structures:
$\mbox{Re}\vec\Delta \parallel \mbox{Im}\vec \Delta$ at $\gamma_2>0$,
and $\mbox{Re}\vec \Delta \perp \mbox {Im} \vec \Delta$ at
$\gamma_2<0$, respectively.
Using the analogy of the spinor condensation of spin-1 bosons,
the former is the polar pairing state and the latter is the
axial pairing state \cite{ohmi1998,ho1998,zhou2003,demler2002}.

\begin{figure}
\centering\epsfig{file=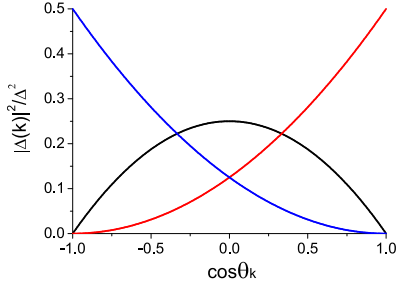,clip=1,width=0.6\linewidth, angle=0}
\caption{The angular distribution of the gap function $|\Delta(\vec k)|^2$
v.s. $\cos\theta_k$ in the helical polar pairing state (the red line)
and the axial pairing state (the black line).
}
\label{fig:gap}
\end{figure}

For the polar pairing state, the order parameter configuration can be
conveniently denoted as $\vec \Delta=e^{i\phi} |\Delta| \hat z$
up to a $U(1)$ phase and $SO(3)$-rotation.
This pairing carries the quantum number $J_z=0$.
The pairing matrix $\Delta^{pl}_{\alpha\beta}=\frac{1}{2}|\Delta|[ k_y \sigma_1 -
k_x \sigma_2) i\sigma_2 ]_{\alpha\beta}$ reads
\bea
\Delta_{\alpha\beta}^{pl}&=&
\frac{1}{2}|\Delta| \left[
\begin{array}{cc}
-(\hat k_y+i \hat k_x)&0  \\
0& \hat k_y-i \hat k_x
\end{array}
\right].
\eea
It equivalents to a superposition of $p_x\mp ip_y$ orbital configurations
for spin-$\uparrow\uparrow$ ($\downarrow\downarrow$) pairs, respectively.
Thus, this pairing state is helical.
It is a unitary pairing state because $\hat\Delta^\dagger \hat\Delta$
is proportional to a $2\times 2$ identity matrix.
The Bogoliubov quasiparticle spectra are degenerate for two different spin
configurations as $E^{pl}_{k,\alpha}=\sqrt{\xi^2_k+|\Delta^{pl}(\vec k)|^2}$
with the anisotropic gap function
$|\Delta^{pl}(\vec k)|^2=\frac{1}{4}|\Delta|^2 \sin^2\theta_k$
depicted in Fig. \ref{fig:gap}.
They exhibit Dirac cones at north and south poles 
with opposite chiralities for two spin configurations.

Similarly, the order parameter configuration in the
axial pairing state can be chosen as $\vec\Delta=
\frac{1}{\sqrt 2}e^{i\phi}|\Delta| (\hat e_x +i\hat e_y)$
up to the symmetry transformation.
This state carries the quantum number of $J_z=1$.
The pairing matrix $\Delta^{ax}_{\alpha\beta}=\frac{1}{2\sqrt 2}|\Delta|
\{[ \hat k_z (\sigma_1 + i\sigma_2)
+ \sigma_z  (\hat k_x+i \hat k_y] i\sigma_2 \}_{\alpha\beta}$
takes the form
\bea
\Delta_{\alpha\beta}^{ax}
&=&\frac{\sqrt 2}{2}|\Delta| \left[
\begin{array}{cc}
\hat k_z&\frac{1}{2} (\hat k_x+i \hat k_y)  \\
\frac{1}{2} (\hat k_x-i \hat k_y)& 0
\end{array}
\right].
\eea
This is a non-unitary pairing state since $\Delta^\dagger \Delta=|\Delta|^2 [
\frac{1}{2}(1+\hat k_z^2) + \hat k_z (\hat k \cdot \vec \sigma)]$.
The Bogoliubov quasiparticle spectra have two non-degenerate branches
with anisotropic dispersion relations as
$E^{ax}_{1,2}(\vec k)=\sqrt{\xi^2_k+|\Delta^{ax}_\pm(\vec k)|^2}$.
The angular gap distribution $|\Delta^{ax}_\pm (\vec k)|^2=
\frac{1}{8} |\Delta|^2 (1\pm \cos\theta_k)^2$ is depicted
in Fig. \ref{fig:gap}.
Each of branch 1 and 2 exhibits one node at north pole and south pole,
respectively.
Around the nodal region, the dispersion simplifies into
$E_{1,2}(\vec k)=\sqrt{v^2_f (k_z\mp k_f)^2+\frac{1}{32} |\Delta|^2
(k_\pp/k_f)^4}$, which is quadratic in the transverse
momentum $k_\pp=\sqrt{k^2_x+k^2_y}$.

\begin{figure}
\centering\epsfig{file=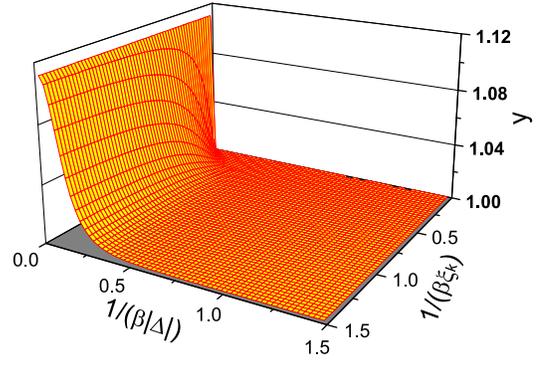,clip=1,width=0.8\linewidth, angle=0}
\caption{The ratio of the angular integrals of the free energy kernels
$y(\frac{1}{\beta |\Delta|},\frac{1}{\beta |\xi|})$, which is
always larger than 1.
This means that the polar pairing is favored at the mean-field level.
}
\label{fig:free}
\end{figure}

At the mean-field level, the helical polar pairing state is more stable
than the axial state.
Actually, this conclusion is not so obvious as in the case of $^3$He-B
phase, where the isotropic gap function is the most stable among all
the possible gap functions \cite{balian1963}.
Here, the gap functions are anisotropic in both the polar and helical
pairing phases.
We need to compare them by calculating their free energies
in Eq. (\ref{eq:free}).
The second term contributes the same to both pairing phases. Thus, the
first term determines the difference in free energies.
Let us define the ratio between angular integrals of the free energy kernels
in Eq. (\ref{eq:free}) of the two phases as
\bea
y(\lambda_1,\lambda_2)= \frac{\int d\Omega_k
~2 \ln \Big [2 \cosh\frac{\beta}{2}
\sqrt{\xi_k^2+|\Delta^{pl}(\vec k)|^2} \Big ]}
{\int d\Omega_k \sum_{\pm}
\ln \Big [2\cosh\frac{\beta}{2} \sqrt{\xi_k^2 + |\Delta^{ax}_\pm(\vec k)|^2}
 \Big ]
}, \ \ \
\eea
where $\lambda_1=\frac{1}{\beta|\Delta|},\lambda_2=\frac{1}{\beta|\xi_k|}$.
$y(\lambda_1,\lambda_2)$ is numerically plotted in Fig. \ref{fig:free}.
For arbitrary values of $\beta$, $\xi_k$, and $|\Delta|$, $y$ is
always larger than 1. Therefore, the polar state is favored more than the
axial state.
This can be understood from the convexity of the nonlinear term in Eq.
(\ref{eq:free}), which favors isotropic angular distributions of
$|\Delta(\vec k)|^2$ \cite{cheng2010}.
Although neither gap function of these two states is
absolutely isotropic as in the $^3$He-B phase, the polar gap
function is more isotropic from Fig. \ref{fig:gap}
and thus is favored.
However, we need to bear in mind that we cannot rule out the possibility
that certain strong coupling effects can stabilize the axial state.
In fact, the $^3$He-A phase can be stabilized under the spin
feedback mechanism \cite{leggett1975}, which is a higher order
effect in terms of interaction strength.

Next we discuss the classification of Goldstone modes and
vortices in these two states.
In the helical polar state,  the remaining symmetries
are $SO_J(2)\times Z_2$ as well as parity and time-reversal (TR), where
$Z_2$ means the combined operation of rotation $\pi$ around any
axis in the $xy$-plane and a flip of the pairing phase by $\pi$.
The Goldstone manifold is
\bea
[SO_J(3)\times U_c(1)]/[SO_J(2)\otimes Z_2]
=[S^2_J\times U_c(1)]/Z_2.
\eea
The Goldstone modes include the phase phonon mode and two branches of
spin-orbital modes.
Vortices in this phase can be classified into the usual
integer vortices in the phase sector and half-quantum vortices
combined with $\pi$-disclination of the orientation of
$\vec \Delta$.
In the axial state, the rotation around $z$-axis generates a
shift of the pairing phase, which can be canceled by a
$U_c(1)$ transformation, thus, the remaining symmetry is $SO_{J_z-\phi}(2)$.
The Goldstone manifold is $S^2\times U_c(1)$.
Only integer vortices exist.

In summary, we have found that the magnetic dipolar interaction provides
a robust mechanism at first order in the interaction strength for a novel
$p$-wave ($L=1$) spin triplet ($S=1$) Cooper pairing state, in which the
total angular momentum of the Cooper pair is $J=1$.
This is a novel pairing pattern which does not appear in $^3$He,
and, to our knowledge, neither in any other condensed matter systems.
These pairing states include the TR invariant helical polar pairing state and
the TR breaking axial pairing state, both of which are
distinct from the familiar $^3$He-$A$ and $B$ phases.

Many interesting questions are open for further exploration, including
the topological properties of these pairing states, vortices,
spin textures, and spectra of collective excitations.
The above theory only applies for spin-$\frac{1}{2}$ systems, in which the
magnetic dipolar interaction is too small.
For the pairing symmetry in a magnetic dipolar system
with a large spin $S$, our preliminary results show that the basic features
of the $J$-triplet pairing remains.
The spins of two fermions are parallel forming $S_{tot}=2S$ with
orbital partial-wave $L=1$, and the total $J=2S$.
In the current experiments in Ref. \cite{lu2012}, the highest attainable density
reaches $4\times 10^{13}$cm$^{-1}$ for $^{161}$Dy atoms with $S=\frac{21}{2}$.
The corresponding dipolar energy is $E_{int}\approx 2$nK and the
Fermi energy for unpolarized gases $E_f\approx 13.6$nK, and
thus $\lambda=E_{int}/E_f\approx 0.15$.
If we use the same formula of $w^{J=1}$ above for an estimation of
the most negative eigenvalue, we arrive at $T_c/T_f\approx 0.06$,
which means that $T_c\approx 0.8$nK.
Although it is still slightly below the lower limit of the
accessible temperature
in current experiments, we expect that further increase of
fermions density, say, in optical lattices will greatly
increase $T_c$.

C. W. thanks J. E. Hirsch for helpful discussions.
Y. L and C. W. are supported by NSF under No. DMR-1105945, and the AFOSR YIP program.


\begin{thebibliography}{10}


\bibitem{lu2010}
Lu, M., Youn, S.~H. and Lev, B.~L.
\newblock Trapping ultracold dysprosium: a highly magnetic gas for dipolar
  physics.
\newblock {\em Phys. Rev. Lett.}{ \bf 104}, 63001 (2010).


\bibitem{youn2010}
Youn, S. H., Lu, M.,  Ray, U. and Lev, B. ~L.
\newblock Dysprosium magneto-optical traps.
\newblock {\em Phys. Rev. A}{ \bf 82}, 043425 (2010).


\bibitem{ospelkaus2008}
Ospelkaus, S. {\it et al}.
\newblock Ultracold polar molecules near quantum degeneracy.
\newblock {\em Faraday Discuss.} { \bf 142}, 351 (2009).

\bibitem{ni2008}
Ni, K.~K. {\it et al}.
\newblock A high phase-space-density gas of polar molecules.
\newblock {\em Science}{ \bf 322}, 231 (2008).

\bibitem{griesmaier2005}
Griesmaier, A., Werner, J., Hensler, S., Stuhler, J. and Pfau, T.
\newblock Bose-einstein condensation of chromium.
\newblock {\em Phys. Rev. Lett.}{ \bf 94}, 160401 (2005).


\bibitem{mcclelland2006}
McClelland, J.~J. and Hanssen, J.~L.
\newblock Laser cooling without repumping: a magneto-optical trap for erbium
  atoms.
\newblock {\em Phys. Rev. Lett.}{ \bf 96}, 143005 (2006).

\bibitem{Aikawa2012}
Aikawa, K. {\it et al}.
\newblock Bose-Einstein Condensation of Erbium.
\newblock {\em }arXiv:1204.1725 { \bf } (2012).

\bibitem{koch2008}
Koch, T., Lahaye, T., Metz, J., Fr{\"o}hlich, B., Griesmaier, A. and Pfau, T.
\newblock Stabilization of a purely dipolar quantum gas against collapse.
\newblock {\em Nat. Phys.}{ \bf 4}, 218--222 (2008).

\bibitem{lahaye2009}
Lahaye, T., Menotti, C., Santos, L., Lewenstein, M. and Pfau, T.
\newblock The physics of dipolar bosonic quantum gases.
\newblock {\em Rep. Prog. Phys.}{ \bf 72}, 126401 (2009).

\bibitem{lahaye2009a}
Lahaye, T., Metz, J., Koch, T., Fr{\"o}hlich, B., Griesmaier, A. and Pfau, T.
\newblock A purely dipolar quantum gas.
\newblock {\em 21st International Conference on Atomic Physics},  160. World
  Scientific, (2009).

\bibitem{menotti2007a}
Menotti, C., Lewenstein, M., Lahaye, T. and Pfau, T.
\newblock Dipolar interaction in ultra-cold atomic gases.
\newblock {\em Dynamics and Thermodynamics of Systems with Long Range
  Interactions: Theory and Experiments}, vol. 970,  332--361,  (2008).



\bibitem{sogo2009}
Sogo, T., He, L., Miyakawa, T., Yi, S., Lu, H. and Pu, H.
\newblock Dynamical properties of dipolar fermi gases.
\newblock {\em New J. Phys.}{ \bf 11}, 055017 (2009).

\bibitem{miyakawa2008}
Miyakawa, T., Sogo, T. and Pu, H.
\newblock Phase-space deformation of a trapped dipolar fermi gas.
\newblock {\em Phys. Rev. A}{ \bf 77}, 061603 (2008).

\bibitem{ronen2010}
Ronen, S. and Bohn, J.~L.
\newblock Zero sound in dipolar fermi gases.
\newblock {\em Phys. Rev. A}{ \bf 81}, 033601 (2010).

\bibitem{chan2010}
Chan, C.~K., Wu, C., Lee, W.~C. and Sarma, S.~D.
\newblock Anisotropic-fermi-liquid theory of ultracold fermionic polar
  molecules: Landau parameters and collective modes.
\newblock {\em Phys. Rev. A}{ \bf 81}, 023602 (2010).

\bibitem{fregoso2009}
Fregoso, B.~M., Sun, K., Fradkin, E. and Lev, B.~L.
\newblock Biaxial nematic phases in ultracold dipolar fermi gases.
\newblock {\em New J. Phys.}{ \bf 11}, 103003 (2009).

\bibitem{lin2010}
Lin, C., Zhao, E. and Liu, W.~V.
\newblock Liquid crystal phases of ultracold dipolar fermions on a lattice.
\newblock {\em Phys. Rev. B}{ \bf 81}, 045115 (2010).

\bibitem{fregoso2010}
Fregoso, B.~M. and Fradkin, E.
\newblock Unconventional magnetism in imbalanced fermi systems with magnetic
  dipolar interactions.
\newblock {\em Phys. Rev. B}{ \bf 81}, 214443 (2010).

\bibitem{fregoso2009a}
Fregoso, B.~M. and Fradkin, E.
\newblock Ferronematic ground state of the dilute dipolar fermi gas.
\newblock {\em Phys. Rev. Lett.}{ \bf 103}, 205301 (2009).

\bibitem{baranov2002}
Baranov, M.~A., Mar\char39{}enko, M.~S., Rychkov, V.~S., and Shlyapnikov, G.~V.
\newblock Superfluid pairing in a polarized dipolar fermi gas.
\newblock {\em Phys. Rev. A}{ \bf 66}, 013606 (2002).

\bibitem{baranov2004}
Baranov, M.~A., Dobrek, L. and Lewenstein, M.
\newblock Superfluidity of trapped dipolar fermi gases.
\newblock {\em Phys. Rev. Lett.}{ \bf 92}, 250403 (2004).

\bibitem{baranov2008a}
Baranov, M.~A.
\newblock Theoretical progress in many-body physics with ultracold dipolar
  gases.
\newblock {\em Physics Reports}{ \bf 464}, 71--111 (2008).

\bibitem{you1999}
You, L. and Marinescu, M.
\newblock Prospects for p-wave paired bardeen-cooper-schrieffer states of
  fermionic atoms.
\newblock {\em Phys. Rev. A}{ \bf 60}, 2324 (1999).

\bibitem{bruun2008}
Bruun, G.~M. and Taylor, E.
\newblock Quantum phases of a two-dimensional dipolar fermi gas.
\newblock {\em Phys. Rev. Lett.}{ \bf 101}, 245301 (2008).


\bibitem{levinsen2011}
Levinsen, J., Cooper, N.~R. and Shlyapnikov, G.~V.
\newblock Topological ${p}_{x}+{\mathit{ip}}_{y}$ superfluid phase of fermionic
polar molecules.
\newblock {\em Phys. Rev. A}{ \bf 84}, 013603 (2011).

\bibitem{potter2010}
Potter, A.~C., Berg, E., Wang, D.~W., Halperin, B.~I. and Demler, E.
\newblock Superfluidity and Dimerization in a Multilayered System of Fermionic
 Polar Molecules.
\newblock {\em Phys. Rev. Lett.}{ \bf 105}, 220406 (2010).

\bibitem{lutchyn2010}
Lutchyn, R.~M., Rossi, E. and Das Sarma, S.
\newblock Spontaneous interlayer superfluidity in bilayer systems of cold polar
	molecules.
\newblock {\em Phys. Rev. A}{ \bf 82}, 061604 (2010).


\bibitem{samokhin2006}
Samokhin, K.~V. and Mar'Enko, M.~S.
\newblock Nonuniform mixed-parity superfluid state in fermi gases.
\newblock {\em Phys. Rev. Lett.}{ \bf 97}, 197003 (2006).

\bibitem{wu2010}
Wu, C. and Hirsch, J.~E.
\newblock Mixed triplet and singlet pairing in ultracold multicomponent fermion
  systems with dipolar interactions.
\newblock {\em Phys. Rev. B}{ \bf 81}, 020508 (2010).


\bibitem{shi2009}
Shi, T., Zhang, J.~N., Sun, C.~P. and Yi, S.
\newblock Singlet and triplet bcs pairs in a gas of two-species fermionic polar
  molecules.
\newblock {\em }arXiv: 0910.4051{ \bf } (2009).


\bibitem{kain2011}
Kain, B. and Ling, H.~Y.
\newblock Singlet and triplet superfluid competition in a mixture of
  two-component fermi and one-component dipolar bose gases.
\newblock {\em Phys. Rev. A}{ \bf 83}, 061603 (2011).


\bibitem{anderson1961}
Anderson, P.~W. and Morel, P.
\newblock Generalized bardeen-cooper-schrieffer states and the proposed
  low-temperature phase of liquid He$^3$.
\newblock {\em Phys. Rev.}{ \bf 123}, 1911 (1961).



\bibitem{balian1963}
Balian, R. and Werthamer, N.~R.
\newblock Superconductivity with pairs in a relative p wave.
\newblock {\em Phys. Rev.}{ \bf 131}, 1553 (1963).

\bibitem{brinkman1974}
Brinkman, W.~F., Serene, J.~W. and Anderson, P.~W.
\newblock Spin-fluctuation stabilization of anisotropic superfluid states.
\newblock {\em Phys. Rev. A}{ \bf 10}, 2386 (1974).

\bibitem{leggett1975}
Leggett, T.
\newblock A theoretical description of the new phases of liquid
$^{3}\mathrm{He}$.
\newblock {\em Rev. Mod. Phys.}{ \bf 47}, 331 (1975).

\bibitem{volovik2009}
Volovik, G.E.
\newblock The Universe in a Helium droplet.
\newblock {\em Oxford University Press}(2009).

\bibitem{ohmi1998}
Ohmi, T. and  Machida, K.
\newblock Bose-Einstein condensation with internal degrees of freedom
in alkali atom gases
\newblock
{\em J. Phys. Soc. Jpn.}{\bf 67}, 1822 (1998).

\bibitem{ho1998}
Ho, T.~L.
\newblock Spinor bose condensates in optical traps.
\newblock {\em Phys. Rev. Lett.}{ \bf 81}, 742--745 (1998).

\bibitem{zhou2003}
Zhou, F.
\newblock Quantum spin nematic states in bose einstein condensates.
\newblock {\em Int. J. Mod. Phys. B}{ \bf 17}, 2643--2698 (2003).

\bibitem{demler2002}
Demler, E. and Zhou, F.
\newblock Spinor bosonic atoms in optical lattices: symmetry breaking and
  fractionalization.
\newblock {\em Phys. Rev. Lett.}{ \bf 88}, 163001 (2002).

\bibitem{lu2012}
Lu, M., Burdick, N. Q., and Lev, B. L.
\newblock Quantum degenerate dipolar Fermi gas.
\newblock{arXiv:1202.4444}.

\bibitem{cheng2010}
Cheng, M., Sun, K., Galitski, V. and Das Sarma, S.
\newblock Stable topological superconductivity in a family of two-dimensional
  fermion models.
\newblock {\em Phys. Rev. B}{ \bf 81}, 024504 (2010).

\end{thebibliography}

\end{document}